\documentclass[twocolumn,aps,pre,superscriptaddress]{revtex4-1}
\usepackage{color}
\usepackage{graphicx}
\usepackage{hyperref}
\usepackage{amsmath,amssymb}
\usepackage{epstopdf}
\usepackage{subcaption}
\begin{document}
	\title{Geometry-induced non-equilibrium phase transition in sandpiles}
	
	\author{M. N. Najafi}
	\affiliation{Department of Physics, University of Mohaghegh Ardabili, P.O. Box 179, Ardabil, Iran}
	\email{morteza.nattagh@gmail.com}
	
	\author{J. Cheraghalizadeh}
	\affiliation{Department of Physics, University of Mohaghegh Ardabili, P.O. Box 179, Ardabil, Iran}
	
	\author{M. Lukovi\'{c}}
	\affiliation{Computational Physics, IfB, ETH Zurich, Stefano-Franscini-Platz 3, CH-8093 Zurich, Switzerland}
	\affiliation{Cellulose and Wood Materials, Empa – Swiss Federal Laboratories for Materials Science and Technology, CH-8600 D\"{u}bendorf  , Switzerland}
	\author{H. J. Herrmann}
	\affiliation{ESPCI, CNRS UMR 7636 - Laboratoire PMMH, 75005 Paris (France)}

	\begin{abstract}
We study the sandpile model on three-dimensional spanning Ising clusters with the temperature $T$ treated as the control parameter. By analyzing the three dimensional avalanches and their two-dimensional projections (which show scale-invariant behavior for all temperatures), we uncover two universality classes with different exponents (an ordinary BTW class, and SOC$_{T=\infty}$), along with a tricritical point (at $T_c$, the critical temperature of the host) between them. The transition between these two criticalities is induced by the transition in the support. The SOC$_{T=\infty}$ universality class is characterized by the exponent of the avalanche size distribution $\tau^{T=\infty}=1.27\pm 0.03$, consistent with the exponent of the size distribution of the Barkhausen avalanches in amorphous ferromagnets (Phys. Rev. L 84, 4705 (2000)). The tricritical point is characterized by its own critical exponents. In addition to the avalanche exponents, some other quantities like the average height, the spanning avalanche probability (SAP) and the average coordination number of the Ising clusters change significantly the behavior at this point, and also exhibit power-law behavior in terms of $\epsilon\equiv \frac{T-T_c}{T_c}$, defining further critical exponents. Importantly the finite size analysis for the activity (number of topplings) per site shows the scaling behavior with exponents $\beta=0.19\pm 0.02$ and $\nu=0.75\pm 0.05$. A similar behavior is also seen for the SAP and the average avalanche height. The fractal dimension of the external perimeter of the two-dimensional projections of avalanches is shown to be robust against $T$ with the numerical value $D_f=1.25\pm 0.01$.
	\end{abstract}

	
	\maketitle
	
In the context of out-of-equilibrium critical phenomena, self-organized critical (SOC) systems have attracted much attention because of their role in a wide range of systems, from finance~\cite{biondo2015modeling} and biological~\cite{jensen1998self} to granular matter~\cite{baule2018edwards}, the brain~\cite{de2006self} and neural networks in general~\cite{makarenkov1991self}. SOC systems are characterized by their avalanche dynamics resulting from slow driving of the system. Vortex avalanche dynamics in type II superconductors~\cite{altshuler2004colloquium}, earthquakes~\cite{gutenberg1956energy}, solar flares~\cite{mendoza2014modelling}, microfracturing processes~\cite{zapperi1997plasticity}, fluid flow in porous media~\cite{lilly1993memory}, phase transition-like behavior of the magnetosphere~\cite{sitnov2000phase}, bursts in filters~\cite{bianchi2018critical}, phase transitions in jammed granular matter~\cite{baule2018edwards}, and avalanches dynamics in the rat cortex~\cite{beggs2003neuronal} are some natural examples for SOC. This large class of natural systems inspired theoretical models with the aim of capturing the dominant internal dynamics that causes the avalanches.\\
Here we find evidence for a new non-equilibrium universality class that is reached by changing the geometry of the underlying graph upon which the model is defined, and analyze the transition point between SOC models. It might be applicable to experiments with spatial flow patterns of transport in heterogeneous porous media~\cite{oswald1997observation}, which involve the toppling of fluid~\cite{najafi2016water}. Another example is the Barkhausen effect in magnetic systems~\cite{spasojevic1996barkhausen}, for which the avalanches have been shown to exhibit scaling behavior with an avalanche size exponent $1.27\pm 0.03$ in amorphous ferromagnets (which constitutes a disordered medium)~\cite{durin2000scaling,mehta2002universal}. From a theoretical perspective there is also an interest in critical phenomena on random geometries.~\cite{gefen1980critical}.\\
We implement the dynamics of the Bak-Tang-Wiesenfeld model (BTW)~\cite{Bak1987Self}, also known as the Abelian sandpile model, on a diluted cubic lattice. This lattice is comprised of sites that are either active (through which sand grains can pass) or inactive (completely impermeable to sand grains), which are labeled by the quenched variable $s$ (called \textit{spin}) that is $+1$ for the active case, and $-1$ for the inactive one. We use the Ising model at finite temperatures ($T$) to obtain the \textit{spin} configuration, which is expressed by the Hamiltonian $H=-J\sum_{<i,j>}s_is_j$, where $s_i$ is the spin on site $i$, $J>0$ the ferromagnetic coupling constant, and $\left\langle i,j\right\rangle $ means that $i$ and $j$ are nearest neighbors. The 3D Ising model undergoes a magnetic phase transition at $T=T_c\approx 4.51$ for the cubic lattice~\cite{preis2009gpu}. Since there is at least one spanning spin cluster \textit{at any temperature} (two sites belong to the same cluster if they are nearest-neighbors and have the same spin) of the 3D Ising model on the cubic lattice, no percolation transition takes place at $T_c$. This can be understood by noting that the critical site percolation threshold for the cubic lattice is around $0.32<0.5$($=$ occupancy probability for $T\rightarrow\infty$ of the Ising model). After constructing an Ising configuration at a given temperature using Monte Carlo, we implement the BTW dynamics on top of the spanning (majority) spin cluster (SSC), i.e. a cluster comprised of spins with the same orientation connecting two opposite boundaries of the lattice. Free boundary conditions are imposed in all directions. In the BTW dynamics, we consider on each site $i$ a height $h_i$ (the number of sand grains) taking initially randomly (independently and uncorrelated) with the same probability one integer from $\left\lbrace 1,...,Z_i\right\rbrace $, in which $Z_i$ is the number of active neighbors of the $i$th site. Then we add a sand grain at a random site $i$, so that $h_i\rightarrow h_i+1$. If this site becomes unstable ($h_i>Z_i$), then a toppling process starts, during which $h_j\rightarrow h_j-\Delta_{i,j}$, where $\Delta_{i,j}=-1$ if $i$ and $j$ are neighbors, $\Delta_{i,j}=Z_i$ if $i=j$, and is zero otherwise. After a site topples, it may cause some neighbors to become unstable and topple, and so on, continuing until no site is unstable anymore. Then another random site is chosen and so on. The average height grows with time, until it reaches a stationary state after which the number of grains that leave the system through the boundary is statistically equal to the number of added ones.
The dynamics can be implemented with either sequential or parallel updating. Criticality of the 3D systems is also manifest in 2D observables, which enables us to apply 2D techniques like critical loop ensemble theory~\cite{dashti2015statistical,najafi2018statistical,najafi2018sandpile,dashti2017bak}. Here we consider three-dimensional (3D) avalanches, as well as their two-dimensional (2D) projections on the horizontal plane.\\
\begin{figure*}
	\begin{subfigure}{0.45\textwidth}\includegraphics[width=\textwidth]{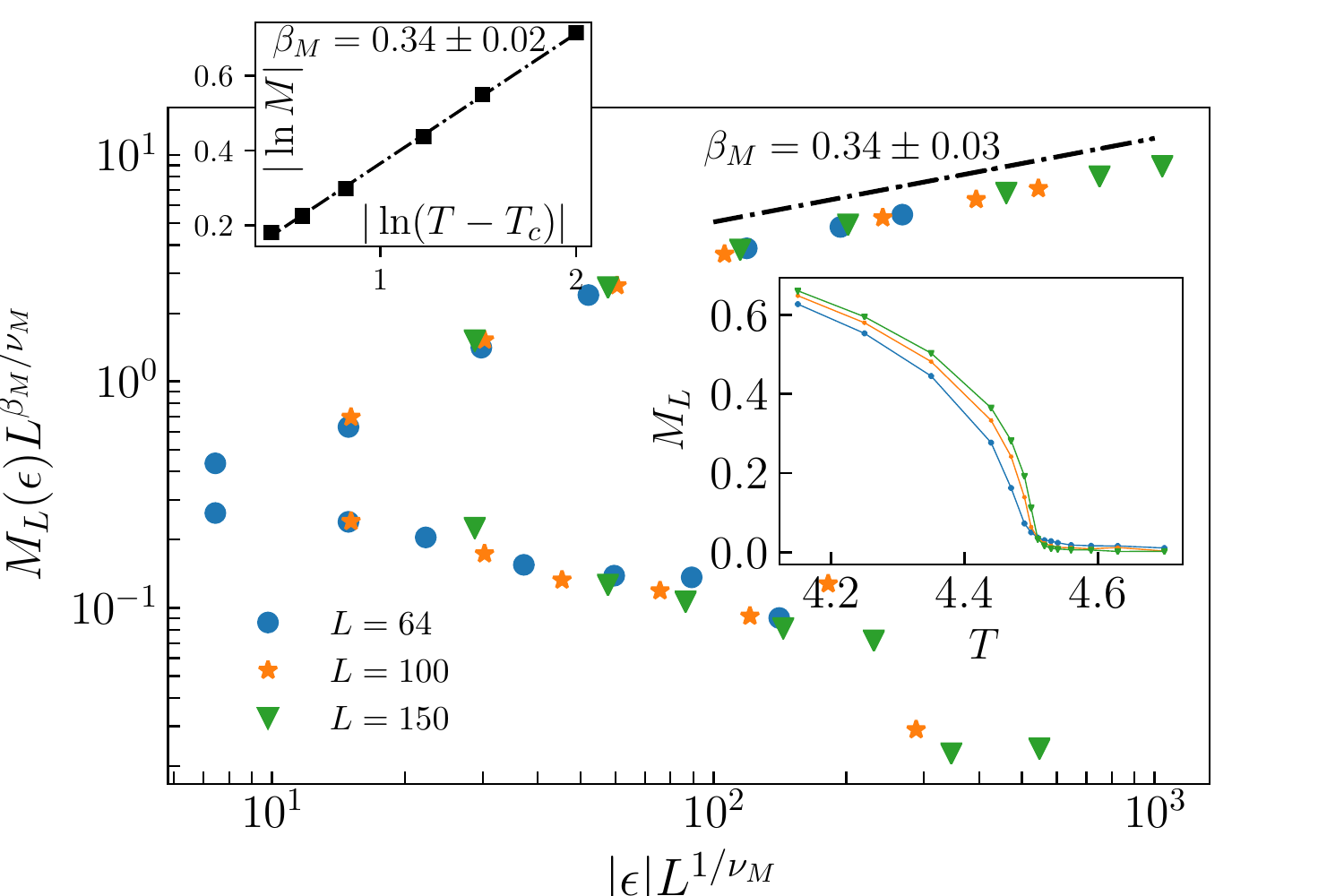}
		\caption{}
		\label{fig:Ising3d}
	\end{subfigure}
	\begin{subfigure}{0.45\textwidth}\includegraphics[width=\textwidth]{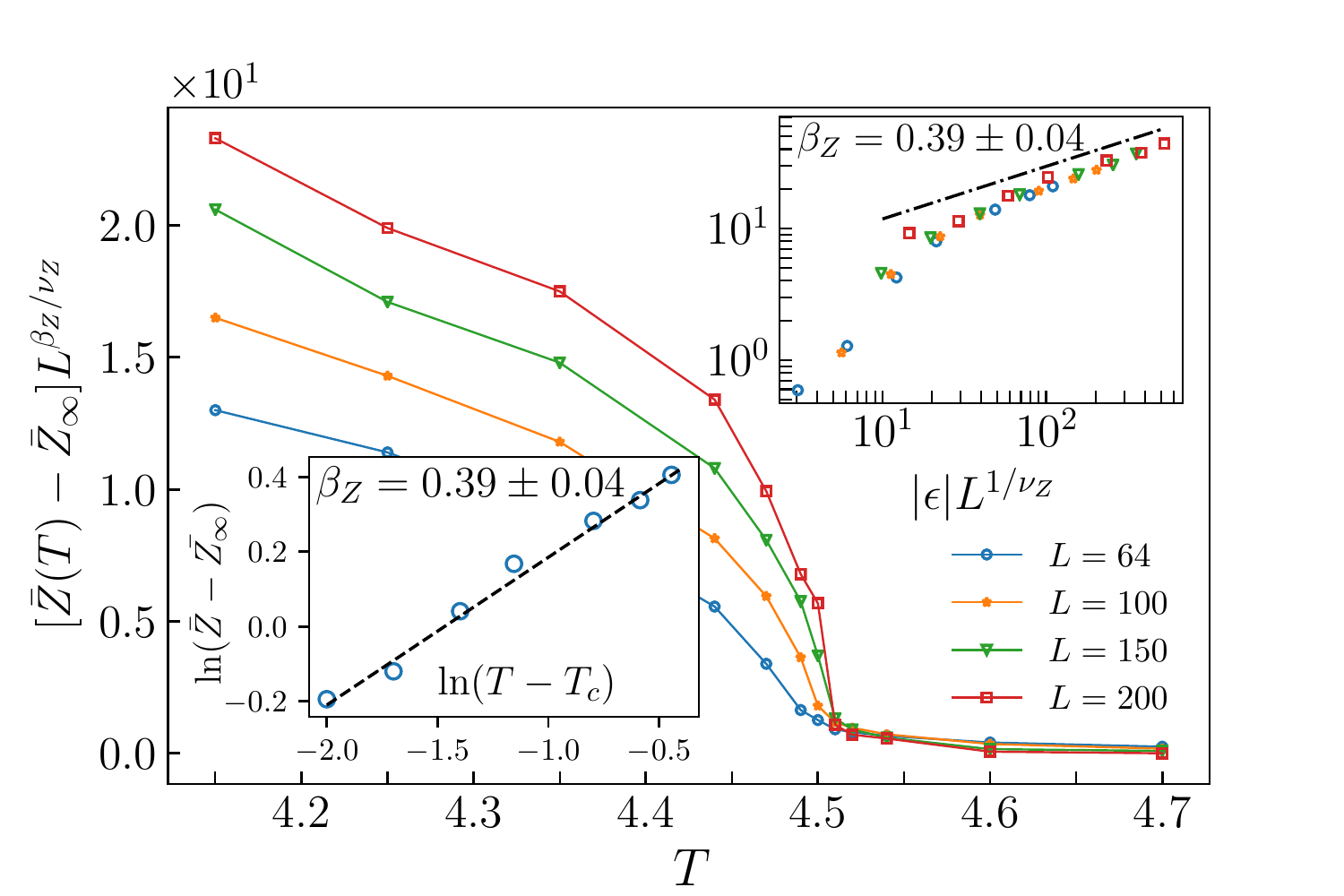}
		\caption{}
		\label{fig:z}
	\end{subfigure}
	\caption{(a) The data collapse of the average magnetization $M_L(\epsilon)$ in terms of $\epsilon\equiv \frac{T-T_c}{T_c}$ with exponents $\beta_M=0.34\pm 0.02$ and $\nu_M=0.63\pm 0.03$. The bare gragh ($M_L$ in terms of $T$) is shown in the lower inset. Top inset: $|\log M|$ in terms of $\log\epsilon$ with the exponent $\beta_M$. (b) The average coordination number $\bar{Z}(T)$ in terms of $T$. It is seen that at $T=T_c$, $\bar{Z}$ changes behavior in a power-law form with the exponents $\beta_Z=0.39\pm 0.04$, $\nu_Z=0.75\pm 0.05$. Top inset: log-log plot of $L^{\beta_Z/\nu_Z}\left( \bar{Z}-\bar{Z}_{\infty}\right)$ in terms of $L^{1/\nu_Z}|\epsilon|$. Lower inset: $\log(\bar{Z}-\bar{Z}_{\infty})$ in terms of $\log\epsilon$.}
	\label{fig:Ising}
\end{figure*}
Before analyzing the avalanches, it is worthy to discuss about the magnetic phase transition of the 3D Ising model at $T=T_c$ which is estimated to be $0.451(3)$ consistent with the previous studies~\cite{preis2009gpu}. The finite size scaling (FSS) analysis of the magnetization per spin ($M$) is presented in Fig.~\ref{fig:Ising} showing that it fulfills the relation $M_L(\epsilon)=L^{-\frac{\beta_M}{\nu_M}}G_M\left(\epsilon L^{\frac{1}{\nu_M}} \right) $ where $\beta_M=0.34\pm 0.01$, $\nu_M=0.63\pm 0.03$, and $G_M$ is a universal function, consistent with previous studies, see Ref.~\cite{talapov1996magnetization}. Therefore a similar behavior change takes place for the fractional number of sites in the majority-spin cluster ($=\frac{1}{2}\left(1+M(T)\right) $). Although the Ising model does not undergo a percolation transition at $T_c$, the average coordination number $\bar{Z}$ of the majority spin cluster dramatically changes behavior at $T_c$. From the Fig.~\ref{fig:z} we observe that $\bar{Z}-\bar{Z}_{\infty}$ decays as a power-law with $\epsilon$, where $\bar{Z}_{\infty}$ is the average coordination number in the $T\rightarrow\infty$ limit. It similarly fulfills the FSS hypothesis, i.e. $\bar{Z}(T)-\bar{Z}_{\infty}=L^{-\frac{\beta_Z}{\nu_Z}}G_Z\left(\epsilon L^{\frac{1}{\nu_Z}}\right)$ with $\beta_Z=0.39\pm 0.04$, $\nu_Z=1.0\pm 0.05$, where $G_Z$ is a universal function. These FSS behaviors are the fingerprint of the second order phase transition at $T=T_c$.\\

\begin{figure}
	\centerline{\includegraphics[scale=.6]{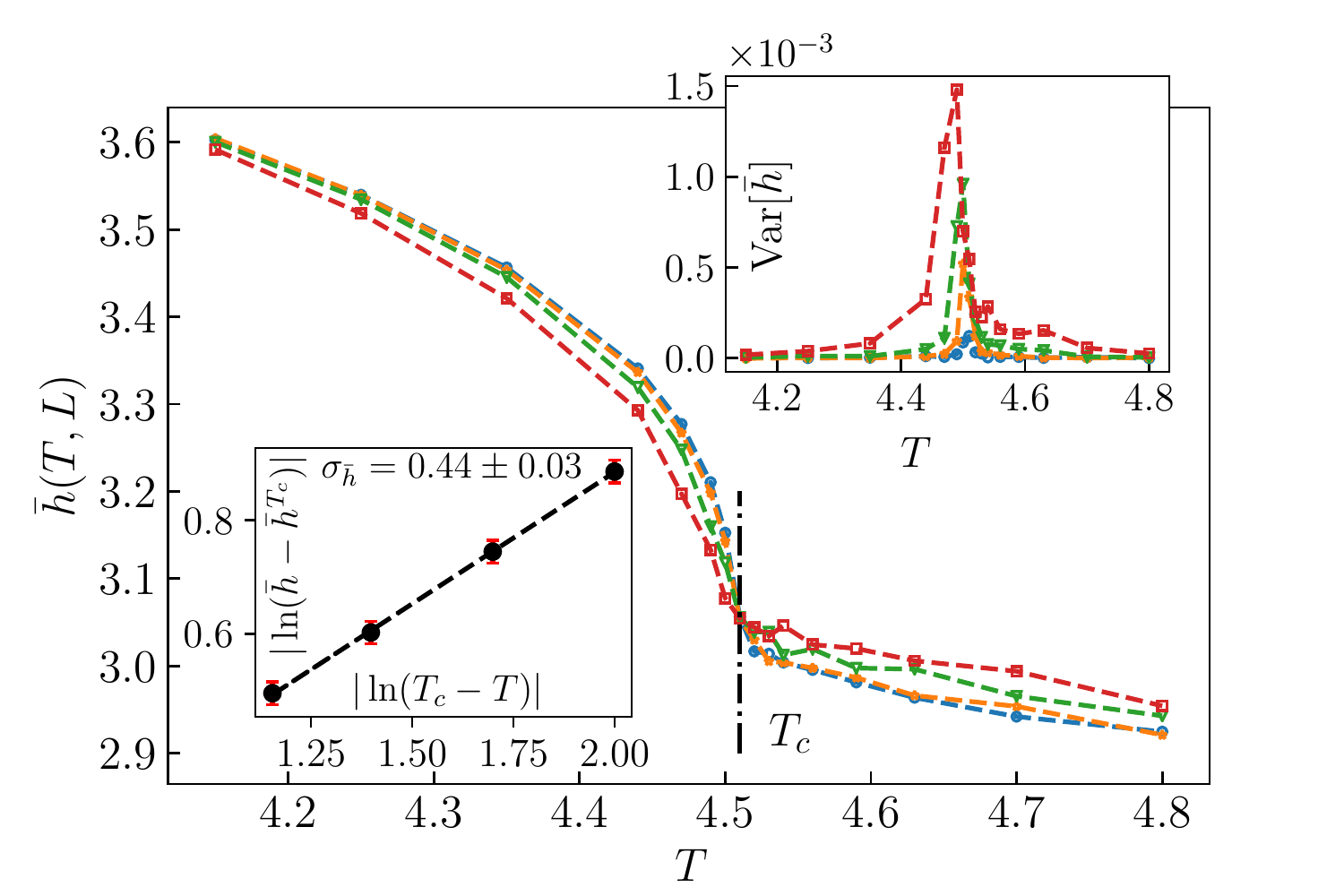}}
	\caption{(a) The average height $\bar{h}(T,L)$ in terms of $T$ for various lattice sizes $L$. It exhibits power-law behavior around $T_c$ as shown in the lower inset: $|\bar{h}(T)-\bar{h}(T_c)|\propto |T-T_c|^{\sigma_{\bar{h}}}$ with an exponent $\sigma_{\bar{h}} = 0.44 \pm 0.03$ for $L=200$. Top inset: the height fluctuation $\text{Var}[\bar{h}]\equiv \left\langle \bar{h}^2\right\rangle -\left\langle \bar{h}\right\rangle^2 $ showing a peak at $T_c$.}
	\label{fig:h}
\end{figure}
Our model undergoes a phase transition at $T=T_c$ separating two different SOC phases, which is induced by the change of the geometry of the support. As the temperature increases, the average height $\bar{h}(T)$ undergoes a substantial change as shown in Fig.~\ref{fig:h}. Close and below $T_c$ it exhibits a power-law decay $\bar{h}(T)-\bar{h}(T_c)\propto |T-T_c|^{\sigma_{\bar{h}}}$ with an exponent $\sigma_h=0.44\pm 0.03$, whereas above this temperature we observe a gentle slow-varying function of $T$. All graphs (for various system sizes $L$) cross each other right at $T=T_c$, at which a pronounced peak arises for the variance of $\bar{h}$, i.e. $\text{Var}[\bar{h}]\equiv \left\langle \bar{h}^2\right\rangle -\left\langle \bar{h}\right\rangle^2 $. \\
Contrary to fixed energy sandpiles, in which $\bar{h}$ acts as a tuning parameter~\cite{rossi2000universality}, in our model the system organizes itself in a critical state, experiencing additional dominant fluctuations (for e.g. $\bar{h}$, see Fig.~\ref{fig:h}) at the transition point. To characterize more precisely the two SOC phases, we analyze the average number of topplings per site (toppling density) in avalanches. We may define as the order parameter $\zeta(T) \equiv f_{\text{perc}}-f(T)$, where $f(T)=\frac{m(T)}{N(T)}$, $m(T)$ being the number of topplings, $N(T)$  the number of sites in the SSC, and $f_{\text{perc}}\equiv f(T=\infty)$.
\begin{figure}
	\centerline{\includegraphics[scale=.6]{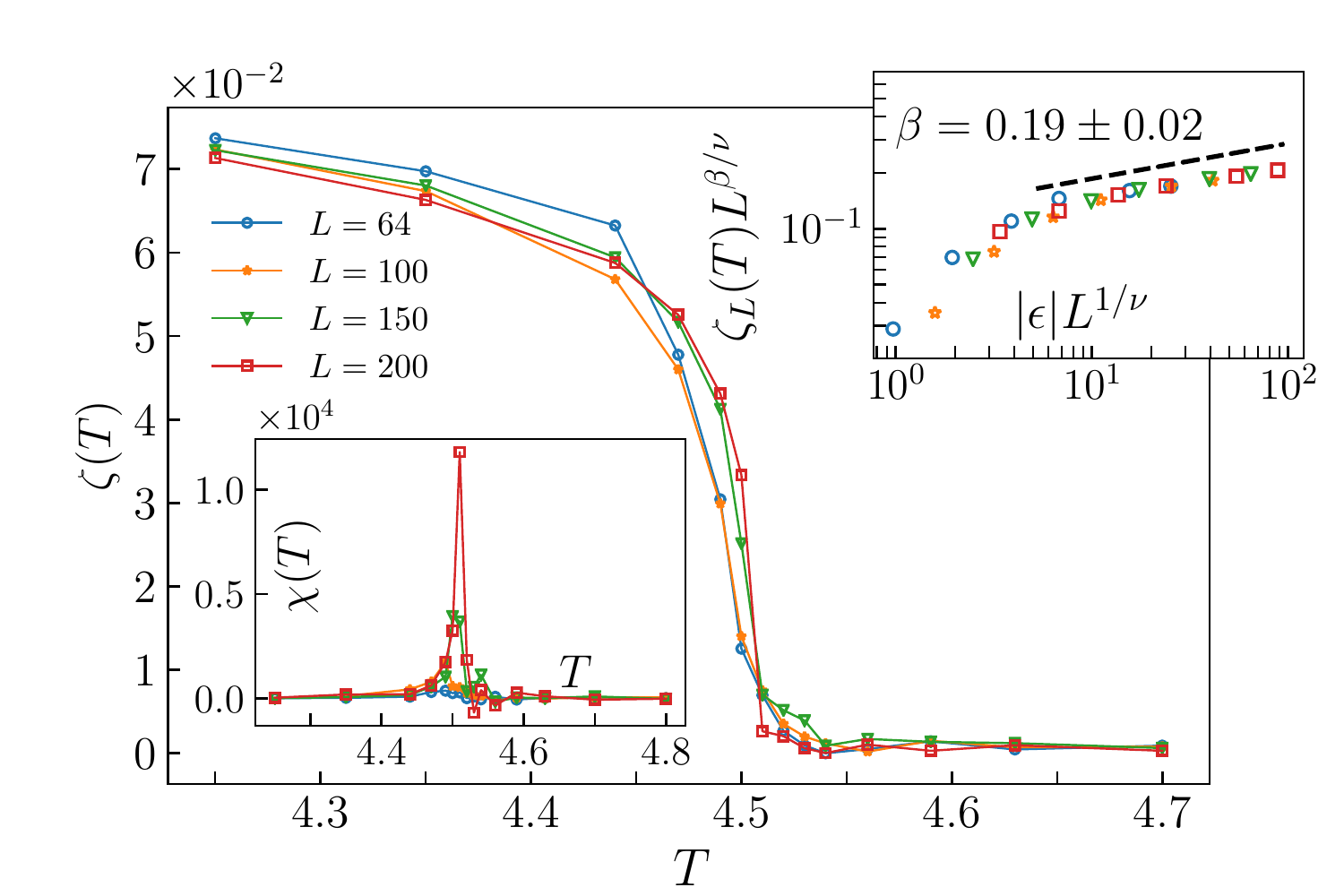}}
	\caption{The order parameter $\zeta(T)$ in terms of $T$ for various lattice sizes. Top inset: data collapse of $\zeta$ with exponents $\beta = 0.19 \pm 0.02$ and $\nu=0.75\pm 0.05$. Lower inset: $\chi(T)\equiv\partial \zeta/\partial T $ in terms of $T$, showing a peak at $T_c$.}
	\label{fig:ave_m}
\end{figure} 
Fig.~\ref{fig:ave_m} reveals that $\zeta(T) $ is approximately zero (is a weak function of $T$) for $T>T_c$, and starts to grow continuously in a power-law fashion when $ T$ is decreased below $T_c$  signaling a phase transition at $T=T_c$, at which $\chi(T)\equiv \frac{\partial \zeta}{\partial T}$ shows a distinct peak. It is notable that $\zeta$ is not precisely zero at $T>T_c$, and has some fluctuations around zero. However these fluctuations diminish as the temperature raises. The FSS relation for $\zeta$ is $\zeta_L(T)=L^{-\beta/\nu}G_{\zeta}\left(\epsilon L^{1/\nu} \right)$ (see upper inset of Fig.~\ref{fig:ave_m}) in which $\epsilon\equiv \frac{T-T_c}{T_c}$, $G_{\zeta}(x)$ is a scaling function with $G_{\zeta}(x)|_{x\rightarrow\infty}\rightarrow x^{\beta}$, and $\beta=0.19\pm 0.02$ and $\nu=0.75\pm 0.05$ are the resulting critical exponents. The case $T\rightarrow\infty$ corresponds to a site percolation cluster with occupation probability $p=\frac{1}{2}$. For $T>T_c$ we will therefore call this phase SOC$_{p=\frac{1}{2}}$. The behavior in the $T<T_c$ region, however, is dominated by $T=0$ i.e. the regular lattice, and therefore we call this phase BTW. The transition between these phases is driven by the connectivity inside the cluster. It becomes clear if we note that $\bar{Z}$ changes character at $T_c$ (Fig.~\ref{fig:z}).
\begin{figure}
	\centerline{\includegraphics[scale=.6]{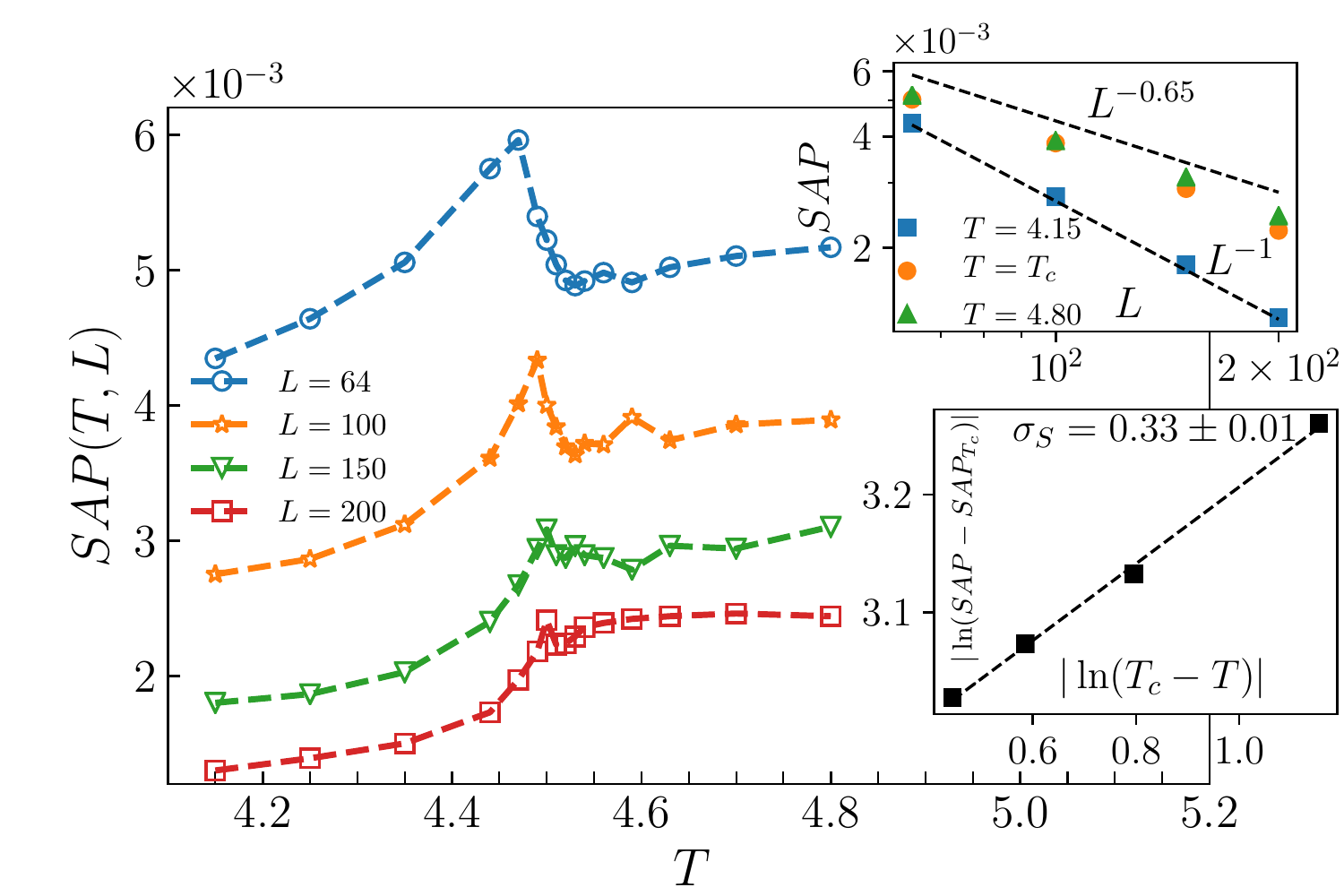}}
	\caption{(Color online): SAP as a function of $T$ and system size $L$, undergoing an abrupt change at $T_c$. The upper inset is the finite size dependence of SAP for $T=4.15<T_c$, $T=T_c$, and $T=4.80>T_c$ from which we see that they all go to zero, for $L\rightarrow \infty$ as power-laws. Lower inset: the power-law dependence of SAP on $T$ for $T < T_c$ close to $T_c$ for $L=200$, defining the exponent $\sigma_S=0.33 \pm 0.01$.}
	\label{fig:SCP}
\end{figure}
The most direct effect of the strong variation of the local coordination number at $T_c$ is the change in the range of avalanches. Let us consider the spanning avalanche probability (SAP), which is the ratio of the number of spanning avalanches (the avalanches that connect two opposite boundaries) to the total number of avalanches, i.e. the probability that an avalanche percolates. This function is a convex monotonic function of $T$ in the $T<T_c$ phase and a slow-varying function at high temperatures ($T>T_c$) (Fig.~\ref{fig:SCP}). It shows a sharp peak at $T=T_c$. We have found that SAP extrapolates to zero for all temperatures, faster for $T<T_c$ than for $T>T_c$. In the upper inset of the Fig.~\ref{fig:SCP} we show this function for three cases: $T=4.15<T_c$, $T=T_c$ and $T=4.80>T_c$ which reveal a power-law decay for all the cases. The increase in SAP when diluting the system is to be expected, since reducing the active channels through which the sand grains can pass increases the range of the corresponding avalanches~\cite{najafi2016bak}. However, this is not true in the vicinity of $T_c$ in which case SAP decreases. This is reminiscent of the sub-diffusive dynamics of random walkers (sand grains here) on the critical percolation cluster ~\cite{argyrakis1984random}. Also the fact that the paths along which the avalanche topples become more tortuous with $T$ (and there are less paths) has a competing effect, and can result in the decrease of the spanning range. Also since the coordination number is reduced, the local firings are less powerful for increasing $T$. In the vicinity of $T_c$, SAP exhibits a power-law behavior in terms of $T$, although it always exhibits power law behavior in terms of $L$. The corresponding exponent $\sigma_S$ is shown in the lower inset of Fig. \ref{fig:SCP} which is defined by $|$SAP$(T)-$SAP$(T_c)|\sim |T-T_c|^{\sigma_S}$ for $T<T_c$. It is shown that $\sigma_S=0.33\pm 0.03$ for $L=200$. \\

\begin{figure}
	\centerline{\includegraphics[scale=.6]{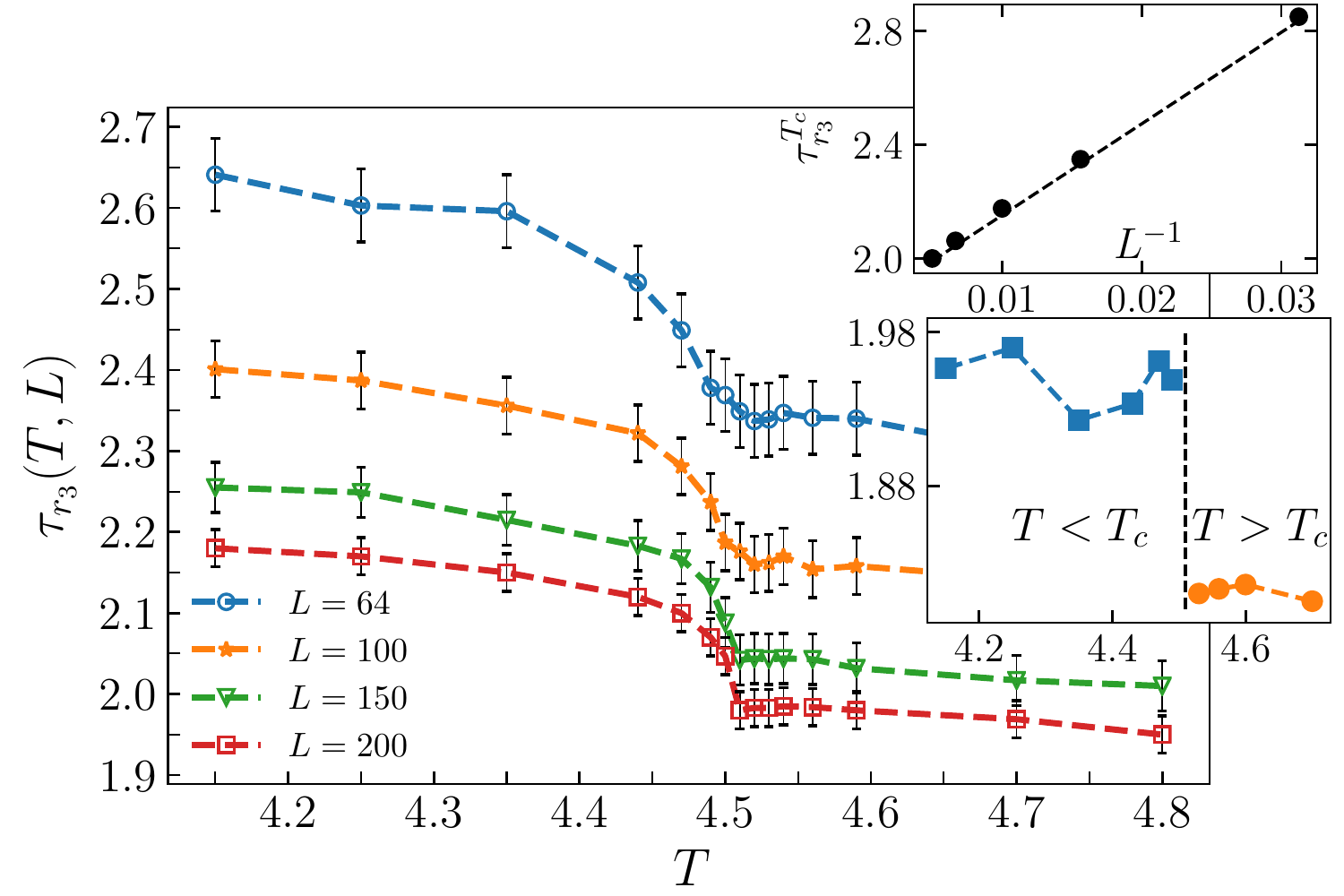}}
	\caption{The critical exponent of the distribution function of gyration radius $\tau_{r_3}$ in terms of $T$. Upper inset: the finite size dependence (linear in terms of $L^{-1}$) of $\tau_{r_3}$ at $T_c$, revealing that $\tau^{T_c}_{r_3}(L\rightarrow\infty)=1.86\pm 0.03$. Lower inset: the same finite size extrapolation of $\tau_{r_3}$ for all temperatures. The exponent undergoes a clear jump at $T_c$.}
	\label{fig:tr3}
\end{figure}
\begin{figure}
	\centerline{\includegraphics[scale=.6]{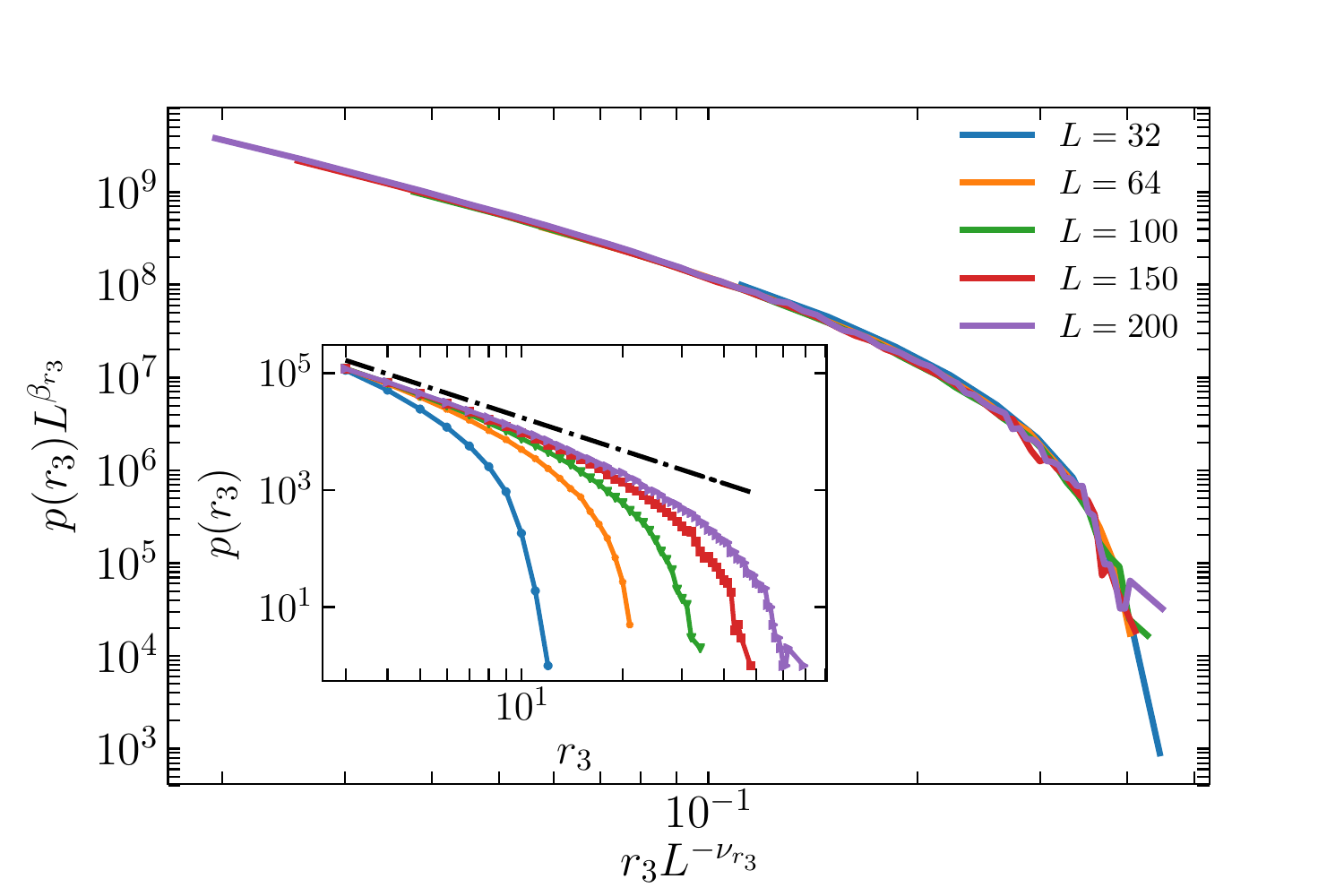}}
	\caption{Data collapse for the distribution function of $r_3$ at $T=T_c$ giving $\beta_{r_3}=1.86\pm 0.03$, $\nu_{r_3}=1.00\pm 0.03$ and $\tau_{r_3}=1.86\pm 0.03$. Inset: distribution of $r_3$ before collapse. The dashed line shows a fit with the slope $\tau_{r_3}=1.83\pm 0.04$.}
	\label{fig:fsc_r3}
\end{figure}

The two phases (ordinary BTW, and SOC$_{\frac{1}{2}}$) and the transition point (SOC$_{T_c}$) can also be characterized in terms of  geometrical quantities. For 3D avalanches: $m_3$ (the mass), $r_3$ (the gyration radius), and $m_2$ (the number of sites in the surface of 3D avalanches), and for 2D projections of avalanches: $r$ (the gyration radius), $l$ (the total length of the external perimeter). For these quantities we study two types of exponents: the exponent of the distribution function $P(x)$, i.e. $\tau_x$ in $P(x)\propto x^{-\tau_x}$, and $\gamma_{xy}$ in $y\propto x^{\gamma_{yx}}$ ($x,y=m_3,r_3,m_2,r$ and $l$). At the transition point all of these exponents show a sharp change(Fig.~\ref{fig:T-Exponents} in SEC.~\ref{App:A}). For example, $\tau_{r_3}$ in the lower inset of Fig.~\ref{fig:tr3} abruptly changes its value from BTW to SOC$_{\frac{1}{2}}$ at $T=T_c$, its value being completely different for $T<T_c$ and $T>T_c$. It is step-like in the limit $L\rightarrow \infty$ which is obtained by linear extrapolation in terms of $1/L$. We observed that for all temperatures $T<T_c$, $\tau_{r_3}$ extrapolates to $1.94\pm 0.04$, and for $T> T_c$ it is $1.76\pm 0.04$. At $T=T_c$, this exponent is $1.86\pm 0.03$ which is different from both values.\\
To be more precise we used also the data collapse technique. Figure~\ref{fig:fsc_r3} shows that the distribution function for $r_3$ (and all other quantities that are analyzed here) fulfills a finite-size scaling relation, i.e. $p(x)=L^{-\beta_x}G(xL^{-\nu_x})$ in which $\beta_x$ and $\nu_x$ are critical exponents and $\tau_x=\beta_x/\nu_x$ that can be used as a check for consistency.  Also note that the cut-off of power-law behavior for the gyration radius (which is commonly considered as the correlation length in sandpile models) as well as the cut-off for the other observables do not practically change with $T$ (see Fig.~\ref{fig:p_r} in~\ref{App:A} for details). The exponents for $r_3$ and $m_3$ for two universality classes ($T=0$ and $T=\infty$) and the transition point $T=T_c$, along with the exponents of $m_2$, $r$, $l$ at $T=T_c$ are gathered in Table~\ref{tab:exponents} (for complete set of exponents see Tables~\ref{tab:exponents} and~\ref{tab:exponents2p} in~\ref{App:A}). We have observed that for $T<T_c$ all the exponents are within the error bars equal to the exponents in the $T=0$ system, whereas for $T>T_c$ they are consistent with $T=\infty$. Most of the exponents are different for two universality classes, and also some exponents for the transition point $T=T_c$ are different from both of them. We note that the size and mass exponent ($\tau_{m_3}^{T=\infty}=1.27\pm 0.03$) in SOC$_{\frac{1}{2}}$ is consistent with the exponent of size distribution of the Barkhausen avalanches in amorphous ferromagnets~\cite{durin2000scaling,mehta2002universal}, which has become important due to its considerable connections with disordered systems and non-equilibrium critical phenomena. Although some authors relate it to the depinning transition~\cite{urbach1995interface,cizeau1997dynamics}, and some others propose a critical point tuned by the disorder in the framework of disordered spin models~\cite{perkovic1995avalanches}, the exact nature of the critical behavior is still debated. Our observation (the similarity between the dynamics of flexible magnetic domain walls in random media, i.e. Barkhausen avalanches and the avalanches of our model) adds the BTW-like avalanches on a diluted lattice to the list of possibilities. This has root in the fact that the maximum pinning force per unit vortex length $f_p^{\text{max}}$ (the defects being realized by 3D percolation) play the role of the threshold for sandpiles above which the particles (vortices) move isotropically toward the neighboring defect site to be pinned, that is reminiscent of the BTW model considered here. Note also that $\tau_{r_3}(T=0)$ was conjectured to be $2$ in Ref~\cite{Lubeck1997BTW}, and also $\tau_{m_3}(T=0)=\frac{4}{3}$, which are consistent with the values reported in Table~\ref{tab:exponents}.\\
\begin{table}
	\begin{tabular}{c | c c c c c }
	\hline quantity & $r_3$ & $m_3$ & $r$ & $l$ & $m_2$ \\
	\hline $\beta(T=0)$ & $1.95(5)$ & $3.71(5)$ & $2.05(5)$ & $2.15(5)$ & $3.85(5)$ \\
	\hline $\beta(T=T_c)$ & $1.86(3)$ & $3.6(1)$ & $1.83(3)$ & $1.96(2)$ & $3.70(5)$ \\
	\hline $\beta(T=\infty)$ & $1.70(5)$ & $3.5(1)$ & $1.80(5)$ & $1.95(3)$ & $3.70(5)$ \\
	\hline
	\hline $\nu(T=0)$ & $1.00(3)$ & $2.74(5)$ & $0.98(3)$ & $1.23(3)$ & $2.75(5)$ \\
	\hline $\nu(T=T_c)$ & $1.00(3)$ & $2.80(5)$ & $1.00(3)$ & $1.20(2)$ & $2.75(5)$ \\
	\hline $\nu(T=\infty)$ & $0.96(4)$ & $2.75(5)$ & $1.05(5)$ & $1.20(2)$ & $2.73(5)$ \\
	\hline
	\hline $\tau(T=0)$ & $1.94(4)$ & $1.32(4)$ & $2.00(5)$ & $1.79(3)$ & $1.38(3)$ \\
	\hline $\tau(T=T_c)$ & $1.86(3)$ & $1.28(2)$ & $1.77(3)$ & $1.62(2)$ & $1.30(2)$ \\
	\hline $\tau(T=\infty)$ & $1.76(4)$ & $1.27(3)$ & $1.75(3)$ & $1.60(3)$ & $1.30(3)$ \\
	\hline
	\hline $\sigma$ & $0.38(4)$ & $0.48(3)$ & $0.52(5)$ & $0.31(5)$ & $0.36(3)$ \\
	\hline
\end{tabular}
	\caption{The critical exponents $\beta$, $\nu$, $\tau$, and $\sigma$ of $r_3$, $m_3$, $r$, $l$ and $m_2$ for two universality classes ($T=0$ and $T=\infty$) and the transition point $T=T_c$. $\beta$ and $\nu$ have been calculated using the data collapse method, and $\tau$ is the linear extrapolation for $L\rightarrow\infty$.}
	\label{tab:exponents}
\end{table}
\begin{table}
	\begin{tabular}{c | c c}
		\hline quantity & $\gamma_{m_3r_3}$ & $D_f\equiv \gamma_{lr}$ \\
		\hline $T=0$ & $2.96(3)$ & $1.25(1)$ \\
		\hline $T=T_c$ & $2.88(3)$ & $1.25(1)$ \\
		\hline $T=\infty$ & $2.84(3)$ & $1.25(1)$ \\
		\hline $\sigma$ & $\gamma_{m_3r_3}$ & $D_f\equiv \gamma_{lr}$ \\
		\hline
	\end{tabular}
	\caption{The fractal dimensions $\gamma_{m_3r_3}$ and $\gamma_{lr}$ for two universality classes, and $T=T_c$. The last row is the $\sigma$ exponent for the two fractal dimensions.}
	\label{tab:exponents2p}
\end{table}

\begin{figure}
	\centerline{\includegraphics[scale=.6]{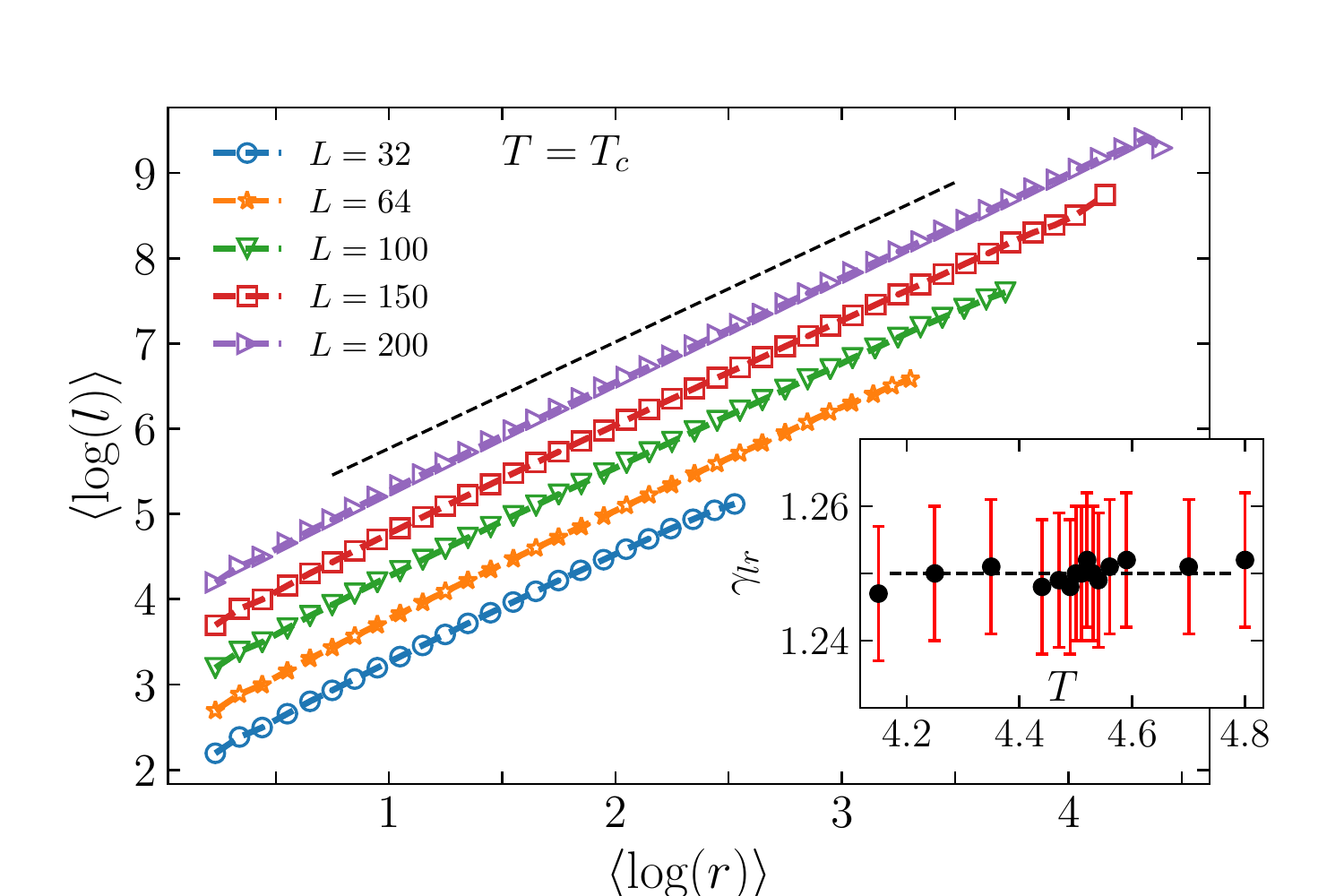}}
	\caption{The ensemble average of $\log l$ in terms of $\log r$ for the external perimeter of avalanches at $T=T_c$. The slope is the fractal dimension $D_f\equiv\gamma_{lr} = 1.25 \pm 0.01$. Inset: the temperature dependence of $D_f$.}
	\label{fig:Df_r1}
\end{figure}
An interesting observation is that the fractal dimension of the avalanche projections ($D_f$ defined by $\left\langle \log l\right\rangle=D_f\left\langle \log r\right\rangle $) is $1.25 \pm 0.01$, consistent with the ordinary 2D BTW model (Fig.~\ref{fig:Df_r1}), and within error bars robust against a change in $T$. A similar phenomenon has been observed in a yet unpublished work in which the fractal dimension of the shadows of clouds (which show some strong similarities with SOC systems) is $1.25$ and is quite robust against the environmental conditions e.g. temperature.\\
Summarizing, we have found a geometry-induced phase transition between two different SOC universality classes at the critical temperature of the Ising model. While at low temperatures the model has ordinary BTW exponents $\tau = 1.34\pm 0.04$ for the avalanche size distribution, at $T_c$ and above $\tau = 1.27 \pm 0.03$. This exponent has been experimentally observed for the size distribution of the Barkhausen avalanches in amorphous ferromagnets~\cite{durin2000scaling,mehta2002universal}, proposing that SOC$_{\frac{1}{2}}$ is a candidate for the universal aspects of Barkhausen noise in magnetic materials. 

\appendix

\section{Other Statistical Observables}
\label{App:A}

\begin{figure*}
	\begin{subfigure}{0.25\textwidth}\includegraphics[width=\textwidth]{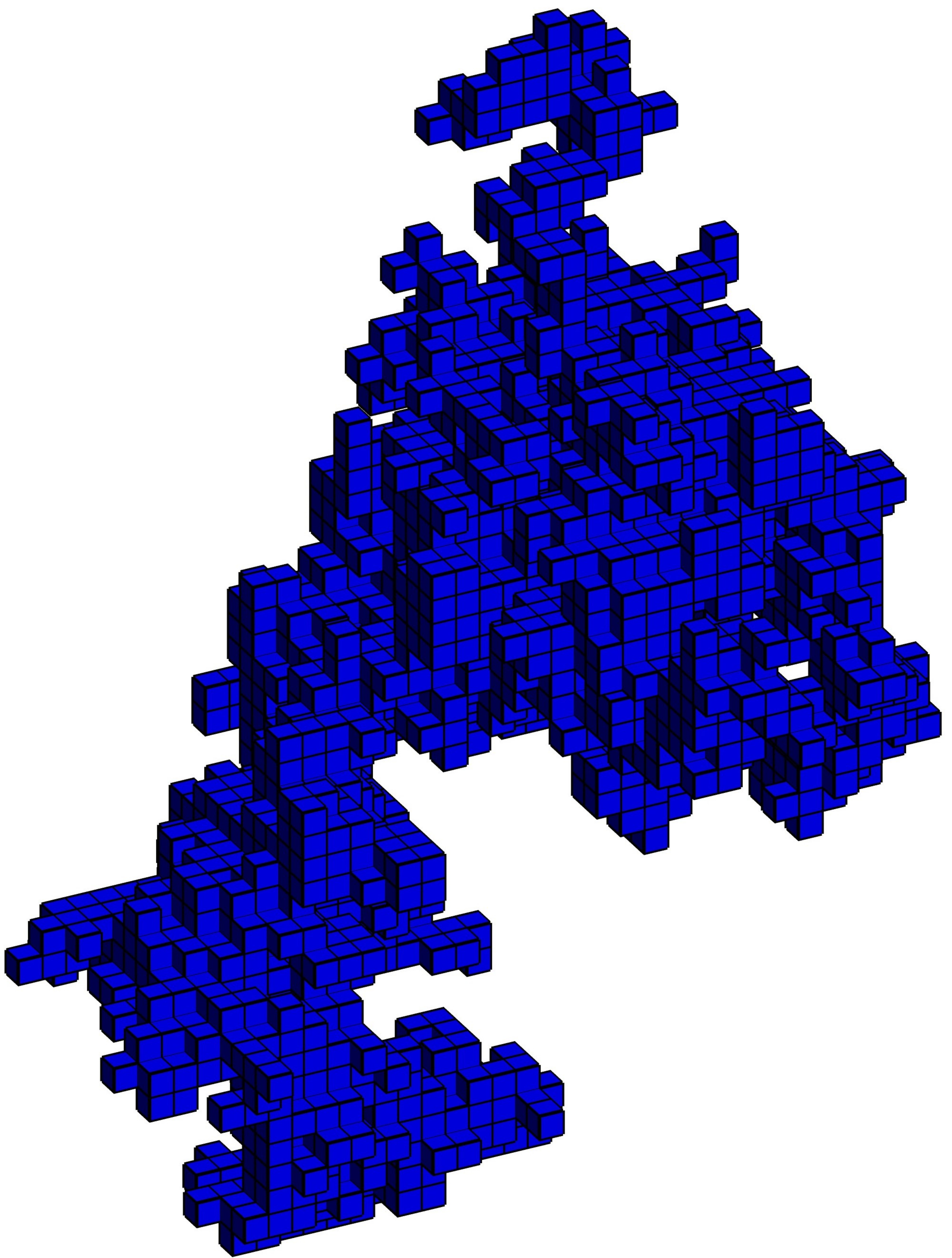}
		\caption{}
		\label{fig:3d-avalanche}
	\end{subfigure}
	\begin{subfigure}{0.25\textwidth}\includegraphics[width=\textwidth]{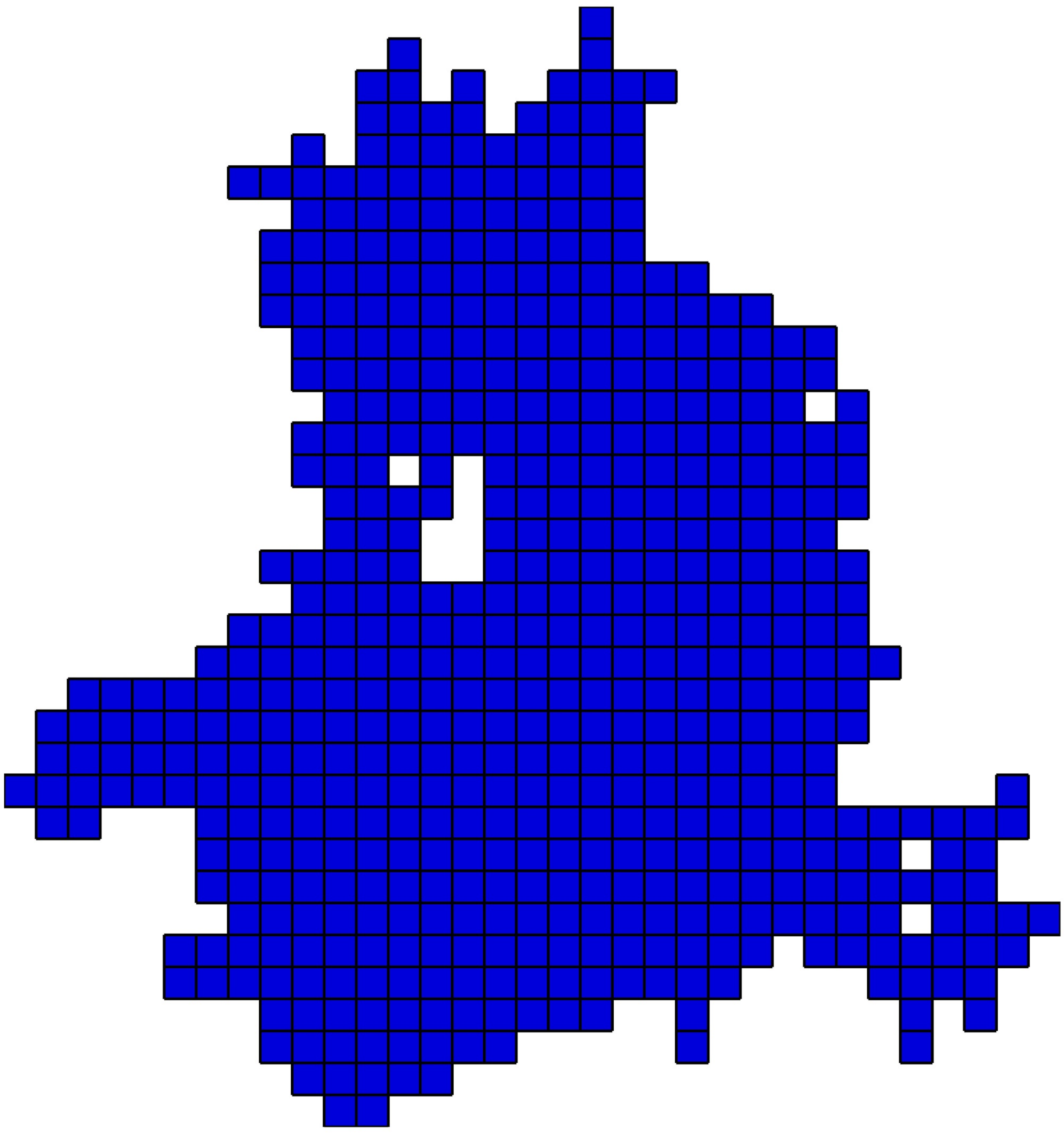}
		\caption{}
		\label{fig:2Dsample}
	\end{subfigure}
	\begin{subfigure}{0.25\textwidth}\includegraphics[width=\textwidth]{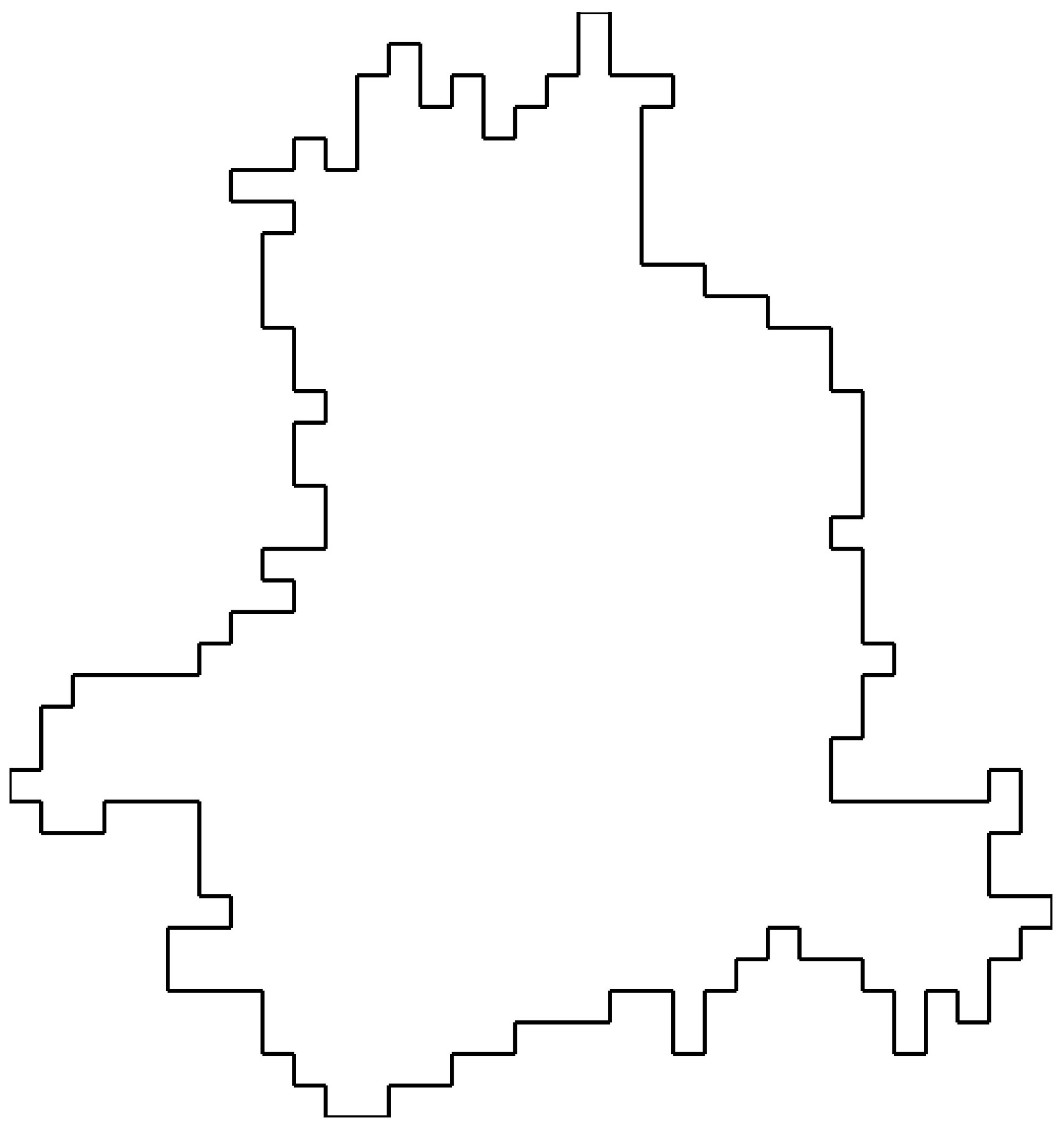}
		\caption{}
		\label{fig:loopSample}
	\end{subfigure}
	\caption{(Color Online) (a). A three-dimensional (3D) BTW sample, and (b) its two-dimensional (2D) projection on the $X-Y$ plane ($Z=0$). (c) External perimeter of 2D shadow.}
	\label{fig:samples}
\end{figure*}

\begin{figure}
	\centerline{\includegraphics[scale=.6]{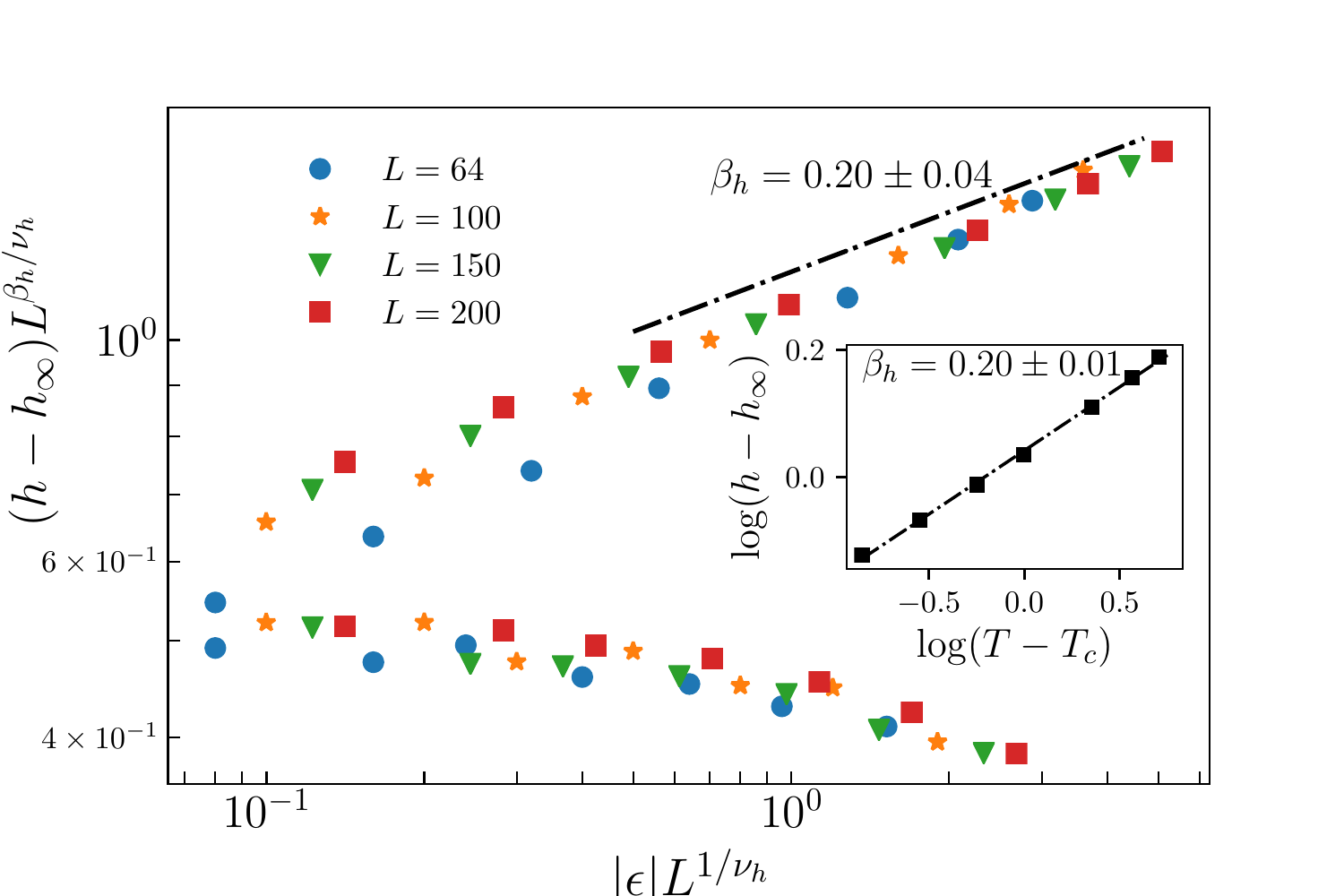}}
	\caption{(Color online): The data collapse for $h_L(T)-h_L(\infty)$ in terms of $\epsilon$. The exponents are $\beta_h=0.2\pm 0.02$, and $\nu_h=2.0\pm 0.1$.}
	\label{fig:h_scale}
\end{figure}

As stated in the paper, we simulate the 3D as well as 2D BTW avalanches on top of the Ising clusters. A sample of the present model (BTW on the cubic Ising lattice) is shown in Fig.~\ref{fig:3d-avalanche}, whose projection is shown in Fig.~\ref{fig:2Dsample}. Also the external perimeter of this sample is shown in Fig.~\ref{fig:loopSample}. An important quantity that changes at this transition point is the average number of toppings per (active) site (the order parameter). The behavior change of the other statistical observables are also interesting at this point. In Fig.~\ref{fig:h} the average height $\bar{h}$ is shown in terms of the host \textit{temperature} $T$, whose changes in behavior is evident at $T=T_c$, with a sharp peak at this temperature. Also $\bar{h}(T)-\bar{h}(T_c)\propto |T-T_c|^{\sigma_{\bar{h}}}$, showing power-law behavior around $T_c$. The data collapse result for $h$ is shown in Fig.~\ref{fig:h_scale}. \\

\begin{figure*}
	\begin{subfigure}{0.45\textwidth}\includegraphics[width=\textwidth]{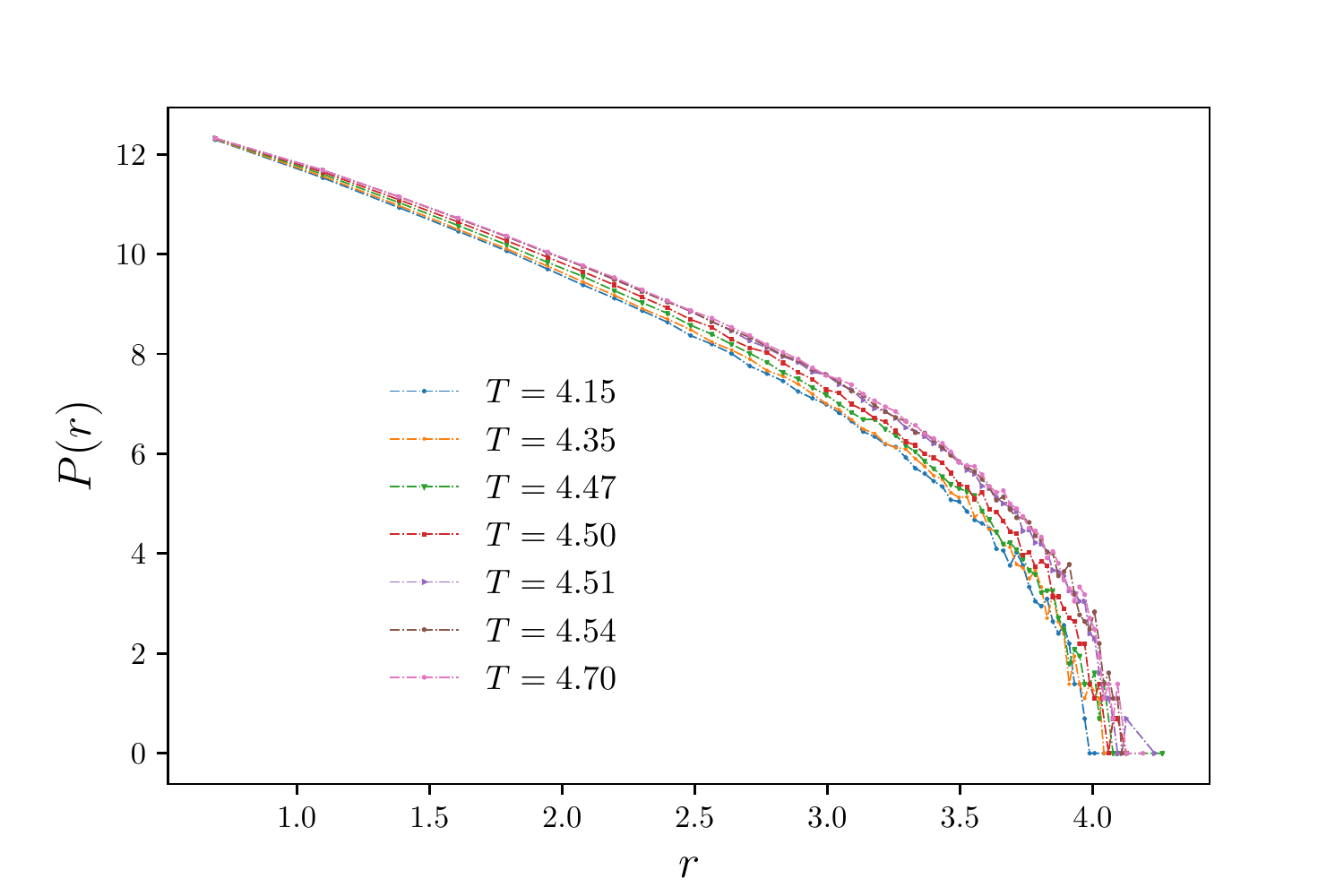}
		\caption{}
		\label{fig:p_r}
	\end{subfigure}
	\begin{subfigure}{0.45\textwidth}\includegraphics[width=\textwidth]{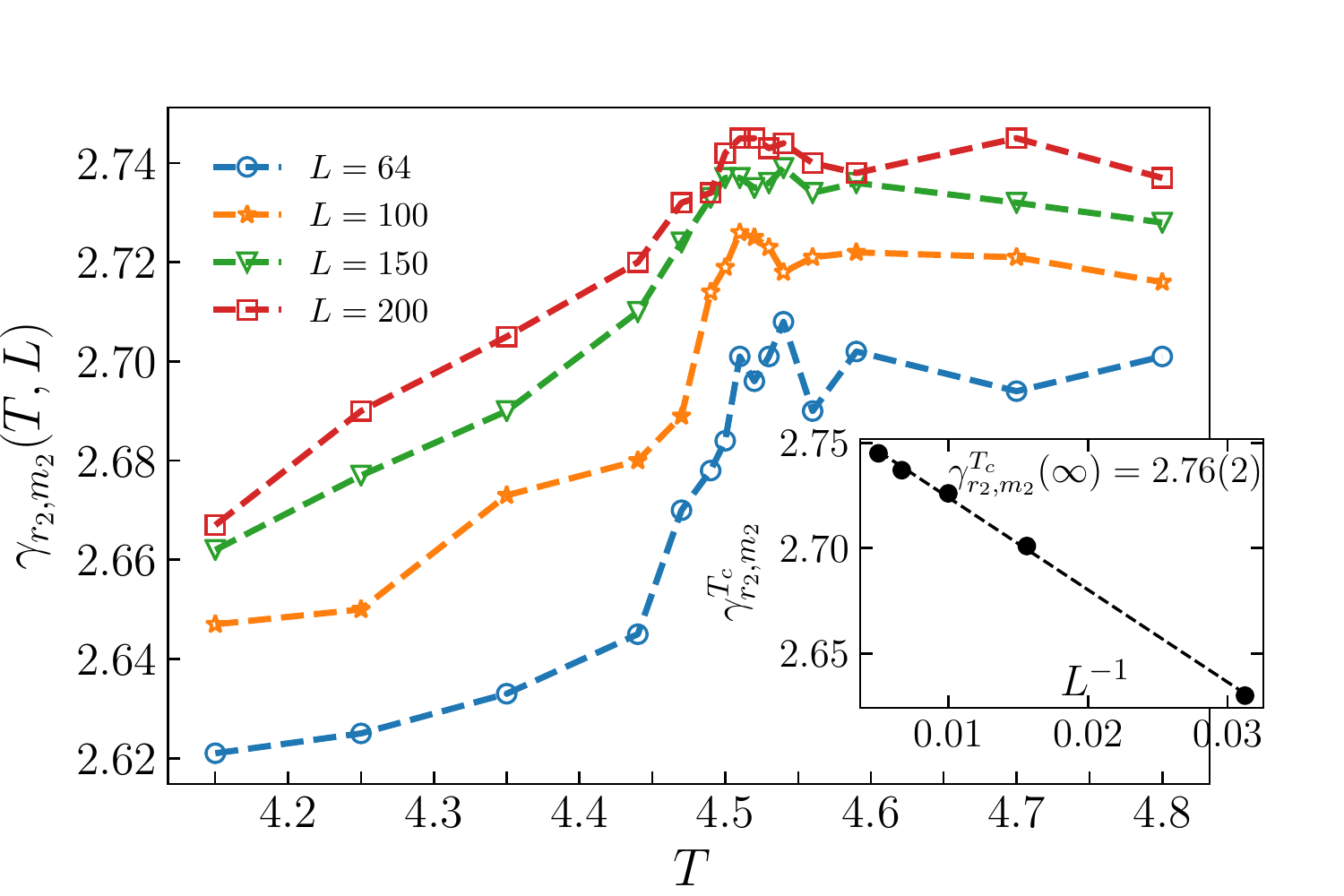}
		\caption{}
		\label{fig:Df_r2}
	\end{subfigure}
	\begin{subfigure}{0.45\textwidth}\includegraphics[width=\textwidth]{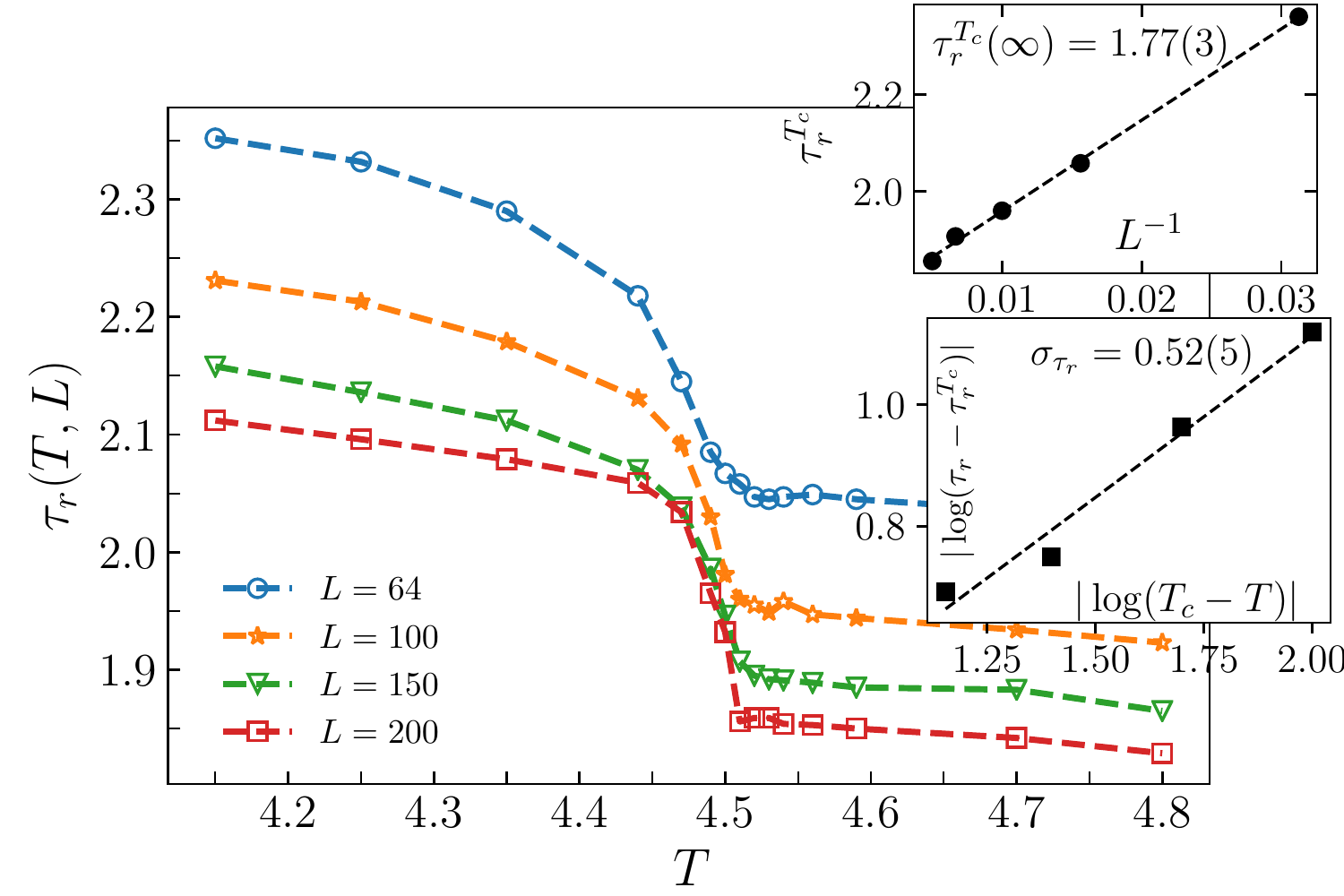}
		\caption{}
		\label{fig:tr1}
	\end{subfigure}
	\begin{subfigure}{0.45\textwidth}\includegraphics[width=\textwidth]{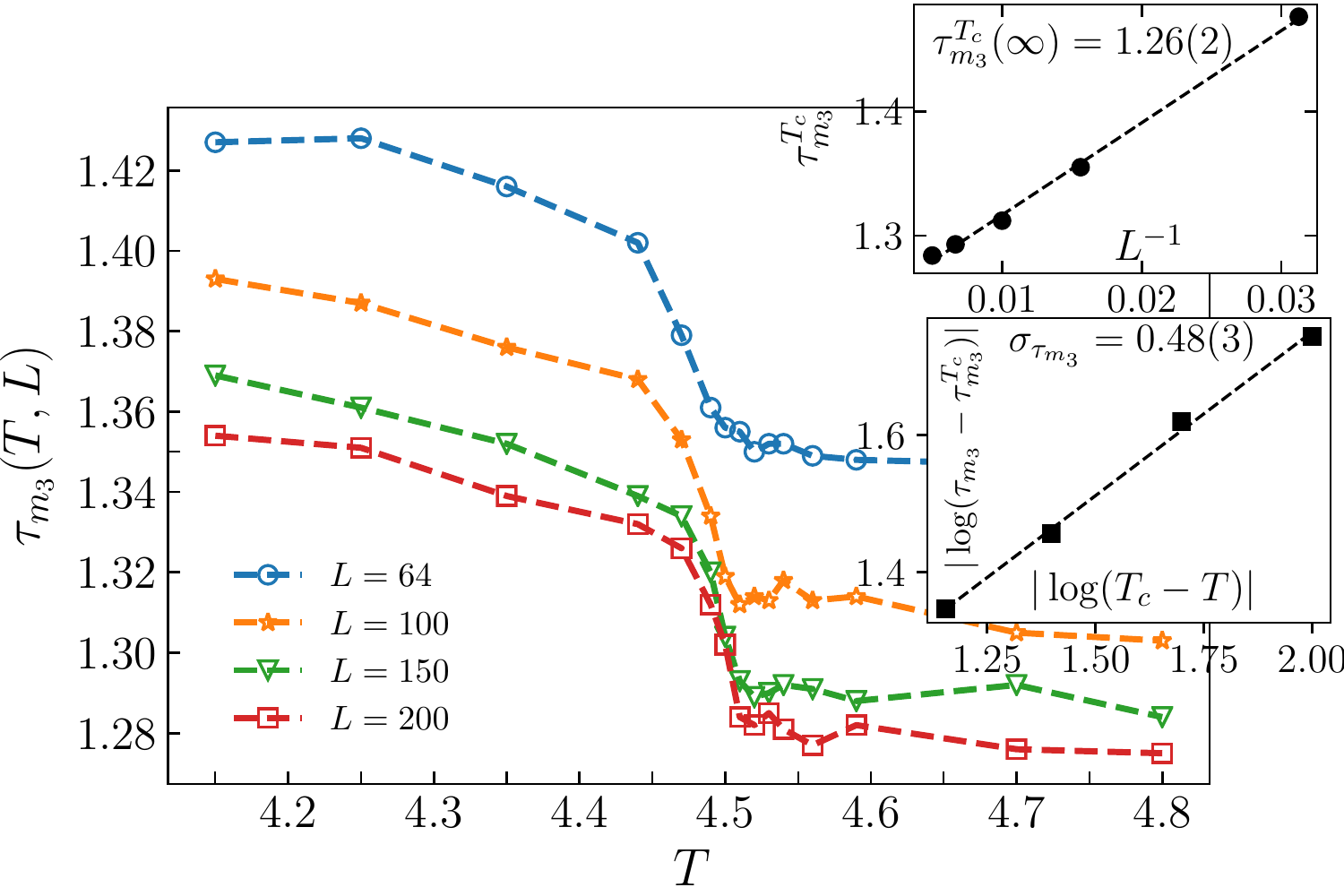}
		\caption{}
		\label{fig:tm3}
	\end{subfigure}
	\begin{subfigure}{0.45\textwidth}\includegraphics[width=\textwidth]{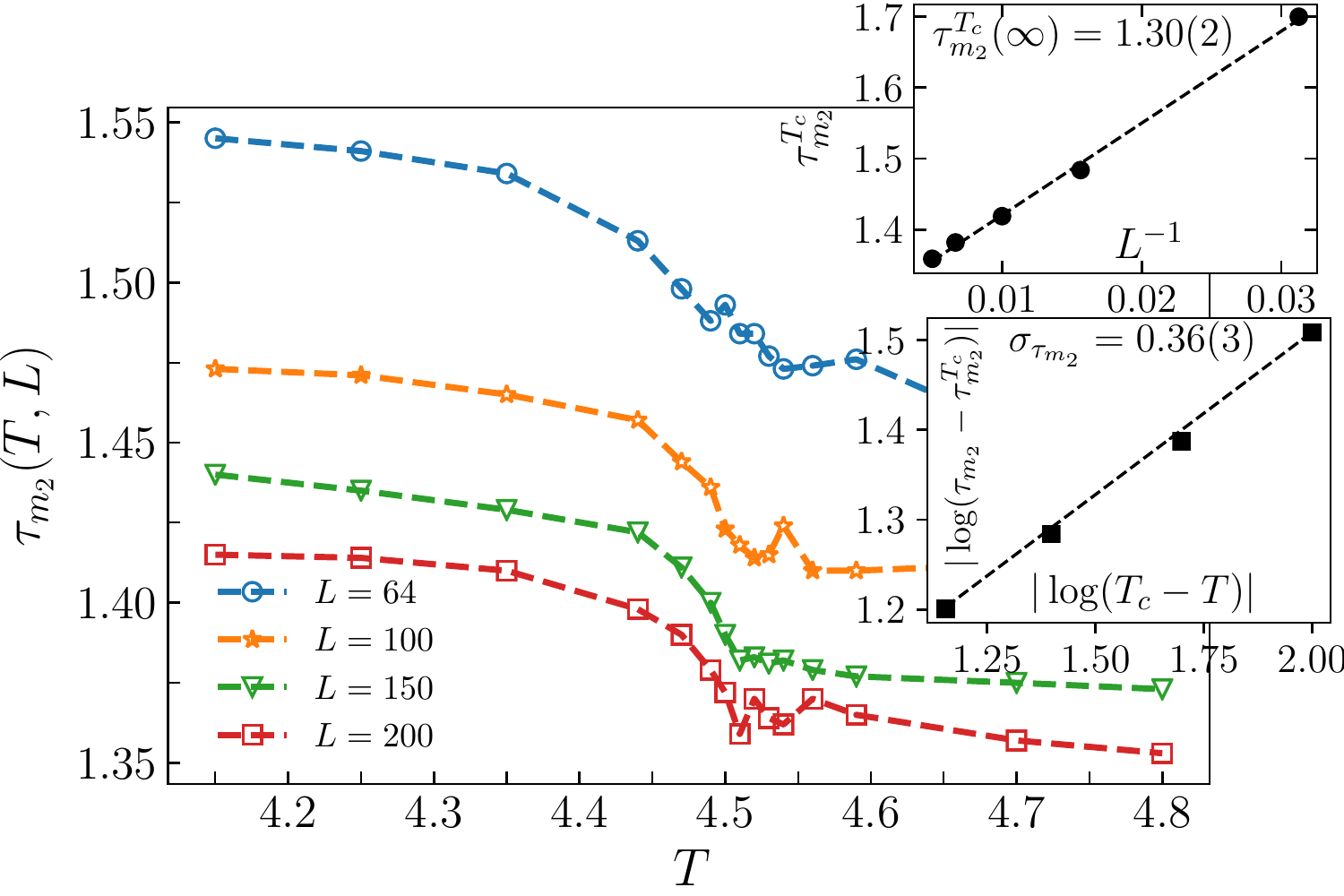}
		\caption{}
		\label{fig:tm2}
	\end{subfigure}
	\begin{subfigure}{0.45\textwidth}\includegraphics[width=\textwidth]{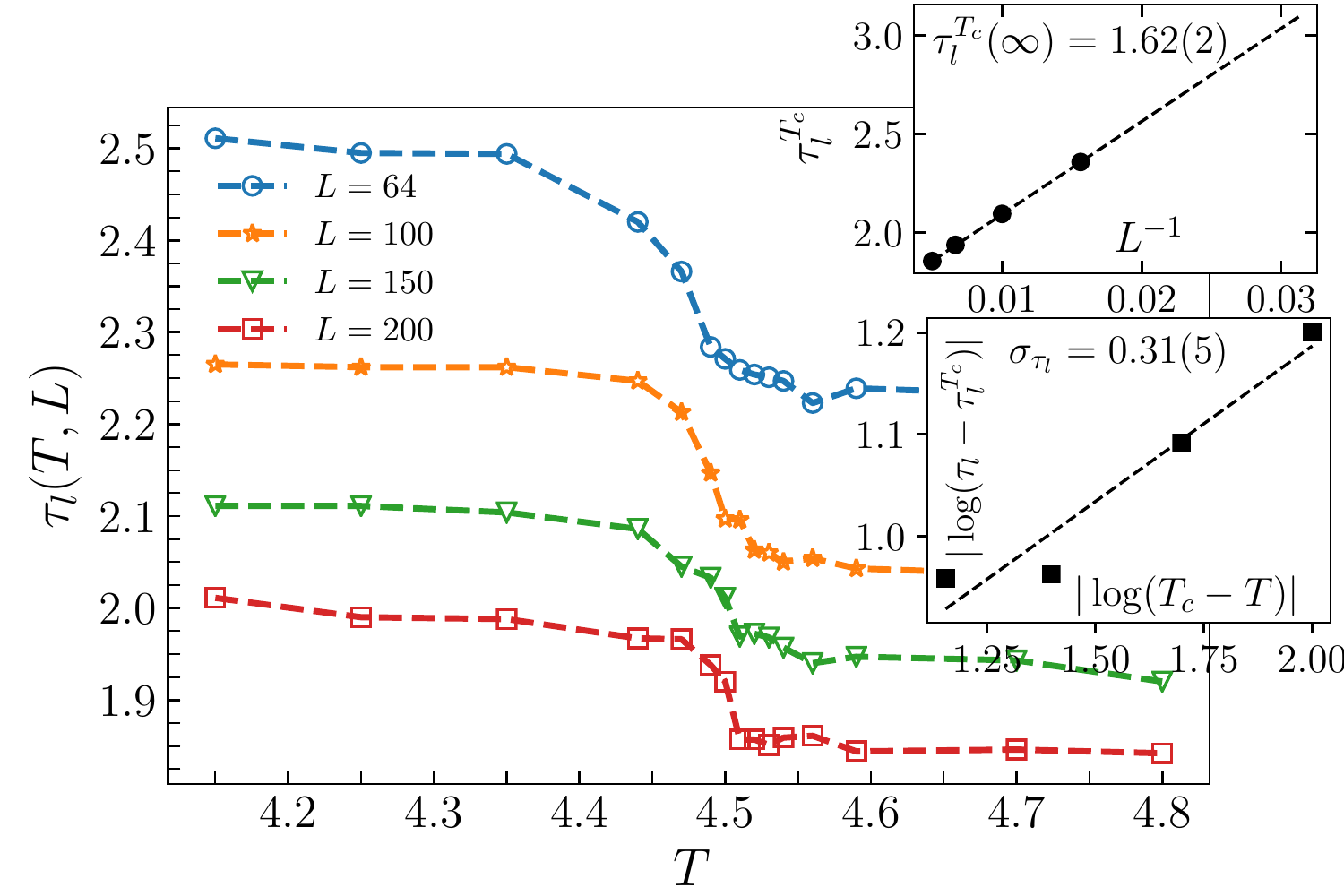}
		\caption{}
		\label{fig:tl}
	\end{subfigure}
	\caption{(Color Online) (a) The log-log plot of the distribution function of the gyration radius for $L=200$. The temperature dependence of (b) $\gamma_{r_2,m_2}$, (c) $\tau_r$, (d) $\tau_{m_3}$, (e) $\tau_{m_2}$, and (f) $\tau_l$. The corresponding exponents, along with finite size analysis are shown in the insets.}
	\label{fig:T-Exponents}
\end{figure*}

\begin{figure*}
	\begin{subfigure}{0.45\textwidth}\includegraphics[width=\textwidth]{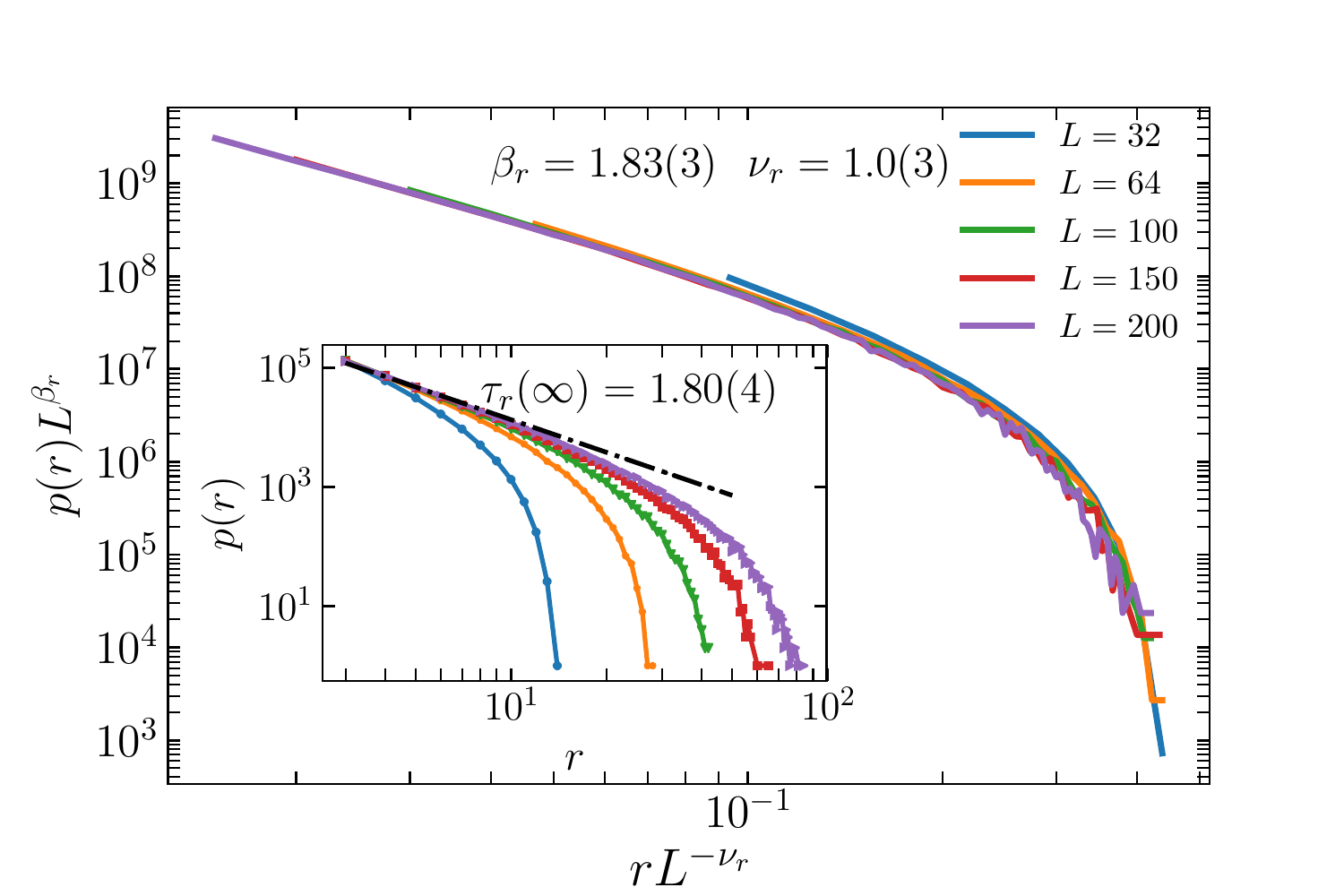}
		\caption{}
		\label{fig:fsc_r}
	\end{subfigure}
	\begin{subfigure}{0.45\textwidth}\includegraphics[width=\textwidth]{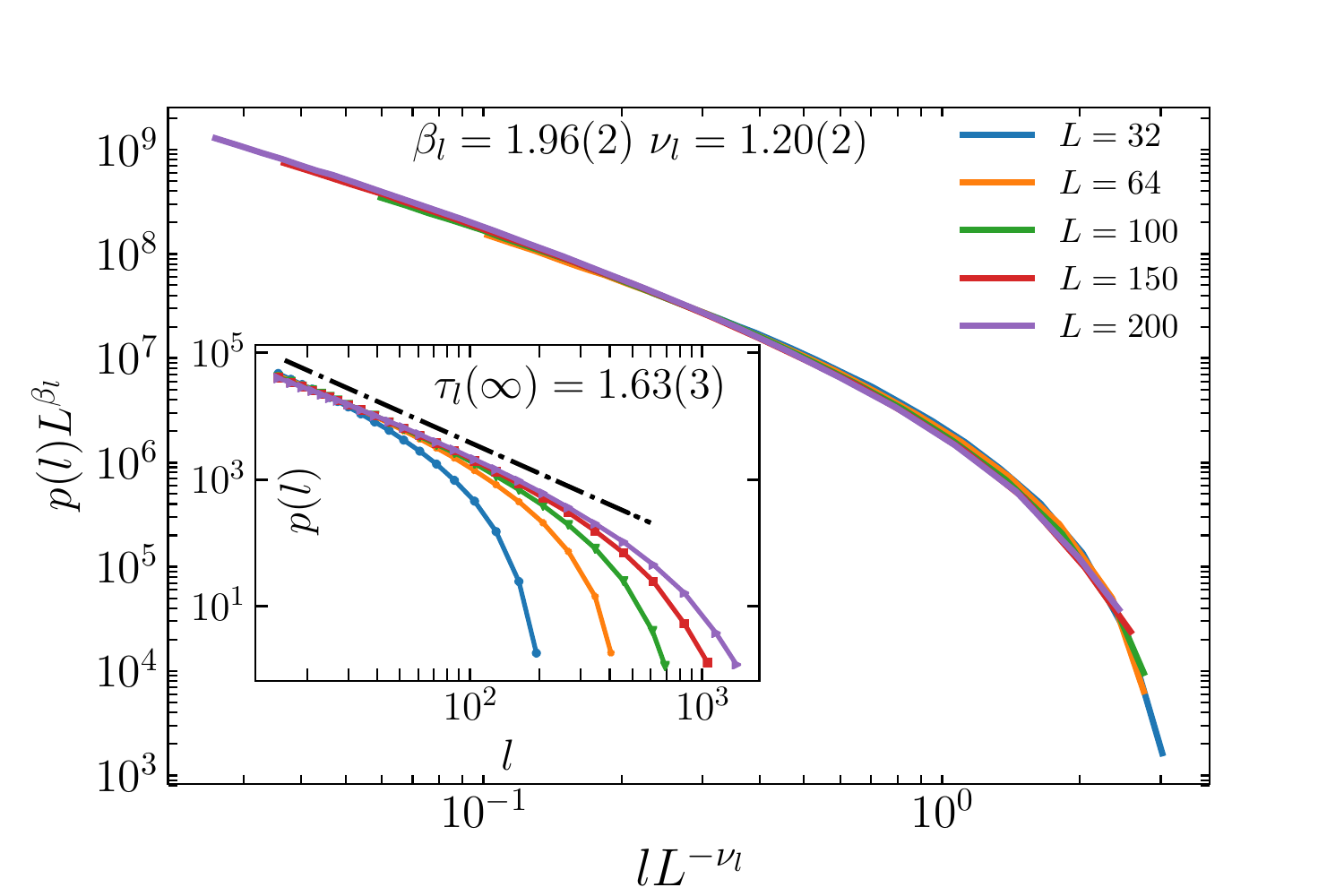}
		\caption{}
		\label{fig:fsc_l1}
	\end{subfigure}
	\caption{(Color Online) Data collapse for distribution function of (a) $r$, and (b) $l$ at $T=T_c$.}
	\label{fig:p_c}
\end{figure*}

We saw that the observables change their behavior at the transition point. An important question then arises concerning the extent of the power-law behaviors in the phase transitions. For conventional phase transitions, the extent of the power-law region reduces as we get away from the critical point. For example, for the Ising model in the vicinity of the critical temperature, this extent reduces with a power of $|T-T_c|$ for all observables. In Fig.~\ref{fig:p_r} we show the log-log plot of the distribution function of the gyration radius for $L=200$. We have observed that $r_{\text{cut}}$ (the cut-off value of $r$ at which the distribution function falls off rapidly) does not run considerably with $T$, in contrast to the conventional second order phase transitions. For example, for two extremes ($T=0$ and $T=\infty$) we know that the power-law behaviors survive already known critical exponents.\\

We have also two type of exponents: the exponent of the distribution function, i.e. $\tau_x$ in $P(x)\propto x^{-\tau_x}$ (in which $x$ is chosen from the above list of statistical quantities), and $\gamma_{xy}$ in $y\propto x^{\gamma_{yx}}$.\\
These quantities ($\gamma_{r_2,m_2},\tau_{r_3},\tau_{r},\tau_{m_3},\tau_{m_2}$ and $\tau_l$) are plotted against $T$ in Fig.~\ref{fig:T-Exponents} for various lattice sizes. An apparent change of behavior is seen at $T=T_c$, below which the exponents are concave (decreasing) functions of $T$, and above which the exponents become (asymptotically) constant. These functions are not analytic at $T=T_c$, i.e. have discontinuity in the first derivative. This discontinuity becomes more significant for larger lattice sizes. Although the exponents are step-like in the thermodynamic limit $L\rightarrow\infty$ (see Fig. 4 in the paper), for finite sizes the exponents show power-law dependence with secondary exponents. These behaviors (concavity for $T>T_c$, change of behavior at $T_c$ have been observed for all other exponents defined above, and power-law behavior around $T_c$ for finite sizes). In each graph both the power-law behavior ($\tau_x(T)-\tau_x(T_c)\sim (T_c-T)^{\sigma_x}$) of the exponent and the finite-size dependence of the exponent at the critical point are shown. The relevant critical exponents (obtained from the power-law behaviors around $T=T_c$) are reported in the graphs. \\

The other thing that we have analyzed is the critical exponents at $T=T_c$, which helps to recognize the universality class of the mentioned phase transition. It is done in~\ref{fig:p_c} in which, using the data collapse technique we extracted the relevant exponents. The analysis has been done for $r_3$, $r$, $l$, revealing that $\tau_{r_3}(T_c)^{L\rightarrow\infty}=1.83(4), \tau_{r}(T_c)^{L\rightarrow\infty}=1.80(4)$ and $ \tau_{l}(T_c)^{L\rightarrow\infty}=1.63(3)$. The exponents for each class ($T=0$ and $T=\infty$), and also for $T=T_c$ are presented in TABLE~\ref{tab:exponents}. It is notable that for $T<T_c$ all exponents are more or less consistent with $T=0$ universality class, whereas for $T>T_c$ they are consistent with $T=\infty$ phase. Some exponents of the transition point $T=T_c$ are different from the ones in the two universality classes.\\
Also the fractal dimension of the 2D projections is shown to be $1.25\pm 0.01$ consistent with ordinary 2D BTW model (Fig.~\ref{fig:Df_r1}). As stated in the paper, we see that this exponent is robust against temperature $T$. The fractal dimensions ($\gamma_{m_3r_3}$ and $\gamma_{lr}\equiv D_f$) can be seen in TABLE~\ref{tab:exponents2p} for $T=0$, $T=T_c$, and $T=\infty$.

\bibliography{refs}

\end{document}